\documentclass[conference,a4paper,twocolumn]{IEEEtran}
\IEEEoverridecommandlockouts

\usepackage{cite}
\usepackage{amsmath,amssymb,amsfonts}
\usepackage{algorithmic}
\usepackage{graphicx}
\usepackage{textcomp}
\usepackage{subfigure}

\def\BibTeX{{\rm B\kern-.05em{\sc i\kern-.025em b}\kern-.08em
    T\kern-.1667em\lower.7ex\hbox{E}\kern-.125emX}}
\begin{document}

\title{Joint Channel Coding and Cooperative Network Coding on PSK Constellations in Wireless Networks\\
}

\author{\IEEEauthorblockN{Elias Benamira$^{1}$, Fatiha Merazka$^{1}$, Gunes Karabulut Kurt$^{2}$}
\IEEEauthorblockA{\textit{$^{1}$LISIC Laboratory, Telecommunications Department} \\
\textit{USTHB University, Algiers, Algeria}\\
\textit{$^{2}$Department of Communications and Electronics Engineering} \\
\textit{Istanbul Technical University, Istanbul, Turkey}\\
ebenamira@usthb.dz, fmerazka@usthb.dz, gkurt@itu.edu.tr}
}

\maketitle

\begin{abstract}
In this paper, we consider the application of Reed-Solomon (RS) channel coding for joint error correction and cooperative network coding on non-binary phase shift keying (PSK) modulated signals. The relay first decodes the RS channel coded messages received each in a time slot from all sources before applying network coding (NC) by the use of bit-level exclusive OR (XOR) operation. The network coded resulting message is then channel encoded before its transmission to the next relay or to the destination according to the network configuration. This scenario shows superior performance in comparison with the case where the relay does not perform channel coding/decoding. For different orders of PSK modulation and different wireless configurations, simulation results demonstrate the improvements resulting from the use of RS channel codes in terms of symbol error rate (SER) versus signal-to-noise ratio (SNR). 
\end{abstract}


\section{Introduction}
The growing interest for reliable wireless networks compels researchers to explore new techniques that exploit network information theory and telecommunications principles, such as the widely investigated network coding and cooperation. This strategy, since its proposal \cite{b1}, attracts a great attention because of its proven efficiency and is already the subject of many research papers including those considering the use of non-binary modulations and joint channel-network coding and decoding. In contrast to traditional communication networks where the nodes can only forward messages individually from different sources, network coding technique allows nodes to process the incoming independent information flows. Namely, a simple linear combination of incoming messages forms a unique message called network coded (NC) message to be forwarded to one or several destinations. This action provides an optimization in network throughput, resources usage and security. With the addition of cooperative communication concept, and thanks to the broadcast nature of wireless networks, a gain in diversity can be achieved as well. 

Exploiting network coding and cooperation principle, the authors in \cite{b2} investigated the use of  low density lattice codes (LDLC) in a  multiple access relay channel (MARC) network formed by two source nodes communicating with one destination via one relay. In two time slots, the two sources broadcast their LDLC coded messages to both the relay and the destination then the relay decodes them and forwards, in the third time slot, their XOR combination to the destination where a joint iterative decoding is performed. With a simulation, using $4-$PAM (Pulse Amplitude Modulation) constellation, they showed that the modulo-addition LDLC outperforms the superposition LDLC and their proposed method provides more diversity and coding gains. In \cite{b3}, channel coded physical layer network coding in a two-way relay scenario was investigated. The authors adopted non-binary $M-$PSK modulation where $M\in\{2,3,4,5\}$. The channel coding schemes used were concatenated RS and convolutional code for $M-$PSK where $M \in \{3,4,5\}$ and LDPC code for BPSK (Binary PSK $i.e.~M=2$). They confirmed by simulation how non-binary constellations $(i.e.$ finite fields $GF(M))$ outperform the binary case in terms of frame error rate (FER) versus signal-to-noise ratio (SNR). 
\begin{figure}[!b]
\centerline{\includegraphics[scale=0.3]{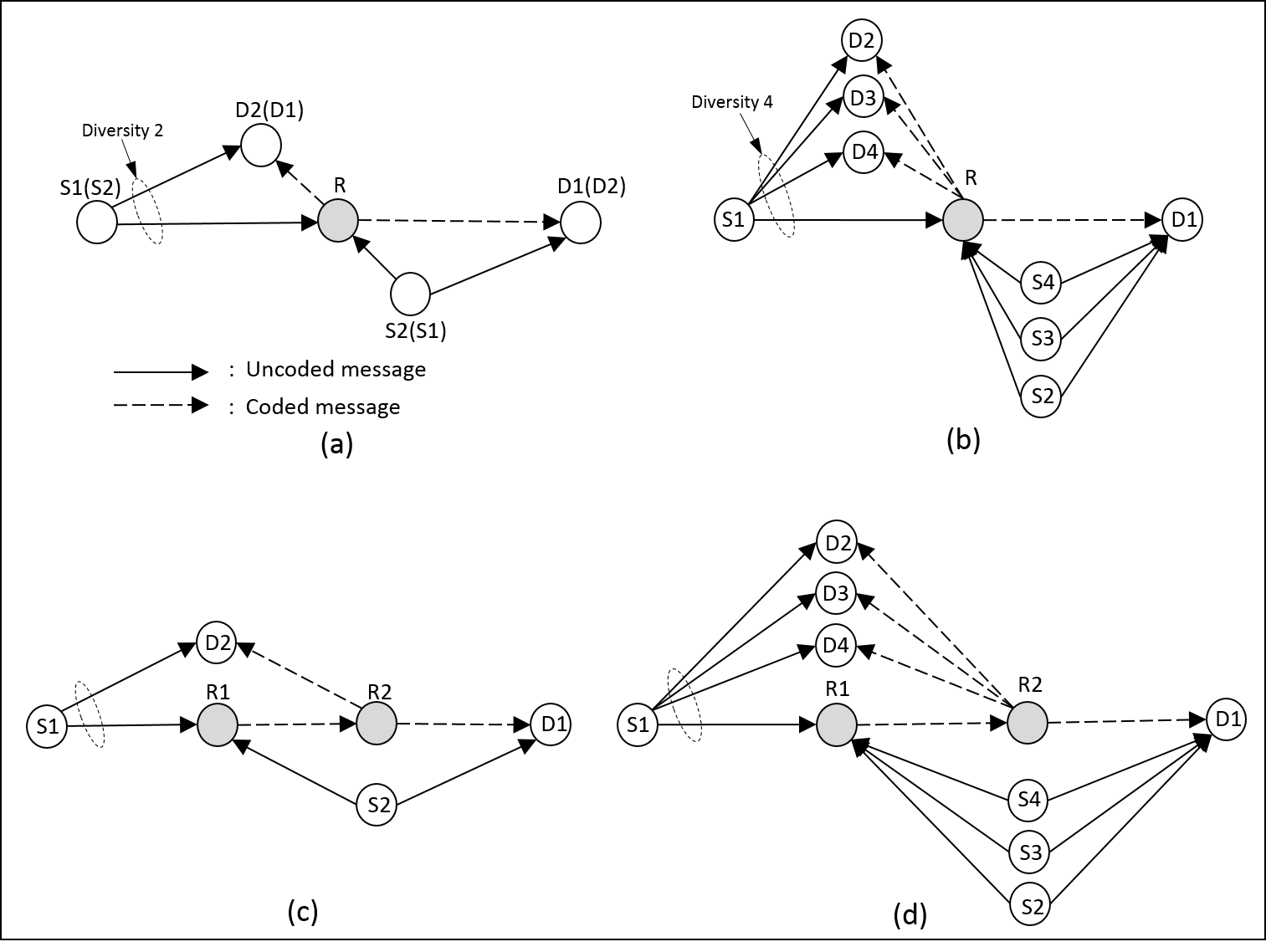}}
\caption{Wireless network topologies considered: (a) X-structure, (b) Extended X-structure, (c) Butterfly network and (d) Extended butterfly network.}
\label{fig}
\end{figure}
The system model used in \cite{b4} contains two source nodes communicating with one sink via one relay and over direct links for cooperation. The authors combined linear network coding with bit interleaved coded modulation  (BICM) scheme using LDPC codes for the time-division 2 users MARC with orthogonal quasi-static fading channels. After simulations with $2^q-ary$ PAM modulation for $q\in\{1,2,3\}$ and $GF(2^q)$ LDPC channel codes, they showed that non-binary LDPC codes outperform the binary case. The authors in \cite{b5} proposed a practical scheme called non-binary joint network-channel decoding (NB-JNCD). In their simulation, they used block fading Rayleigh channels $i.e.$ a constant fading coefficient for each transmitted packet of 1000 LDPC coded symbols (coding rate of $0.8$) over $GF(2^4)$ with $16-$QAM (Quadrature AM) modulation.

The rest of the paper is organized as follows. Section II describes the system model and the equations handling the different scenarios in the wireless network configurations considered. In Section III, we first present the assumptions and parameters of our simulations and then we expose the numerical results in graphs showing the evolution of SER versus SNR for all scenarios and all network topologies. Finally, we conclude our work in Section IV.

\section{System Model}

In this paper, we aim to highlight the advantages of combining NC and cooperation (NCC) with RS channel codes on some widely used wireless topologies for higher orders of PSK constellations. In \cite{b6}, the authors exposed in details many sophisticated channel coding schemes such as RS codes, LDPC codes and Turbo codes. We opt for Reed-Solomon codes because they are well-suited for non-binary symbols transmission over wireless channels. RS codes are  non-binary cyclic codes with symbols made up of $q-$bit sequences, where $q$ is any positive integer having a value greater than 2 which makes them suitable for coding symbols in a finite field $GF(M)=GF(2^q)$, $i.e.$, taking values in $\{0,1,...,M-1\}$. The RS coded values of elements in $GF(M)$ are themselves elements of the same field, $i.e.$, if for instance $M=8$ then we have: $\forall s \in GF(8), \Gamma(s) \in GF(8)$ where $\Gamma$ represents the RS coding operator. In the studied networks, the MAC phase, where $N$ source nodes transmit to the relay, lasts $N$ time slots (straightforward network coding). The broadcast phase, where the relay broadcasts only one network coded symbol instead of $N$, is the key to reducing the amount of information transmitted through the network and hence to throughput increase. Furthermore, the use of high order modulation provides high bandwidth efficiency. In this work, we modulate symbols in $GF(8)$, $GF(16)$  and $GF(32)$ using $8-ary$, $16-ary$ and $32-ary$ PSK constellations respectively. All channels are independent (correlation coefficient $\rho=0$) and subject to AWGN and fast Rayleigh attenuation. Perfect channel state information (CSI) is assumed available at each receiver node from its direct transmitter node. All source nodes have channel coding/decoding capabilities, all relays in charge of NC procedure can also perform channel coding/decoding and finally all destination nodes are equipped with channel and network decoders. We first show the advantages of NCC scheme in comparison with the direct path with and without channel coding in terms of throughput, transmit power and diversity gains, then demonstrate how RS codes can reduce considerably the SER and finally we compare the performance of the two following schemes combining RS channel coding and NCC:
\begin{enumerate}
\item \textbf{Scheme 1}: The relay receiving the RS coded symbols from the sources performs the network coding on them without channel decoding. The destinations apply network decoding before RS decoding. In other words, the relays in charge of network coding are not equipped with channel coders and decoders.

\item \textbf{Scheme 2}: The relay receiving the RS coded symbols from the sources first decodes each one of them before applying network coding and then re-encodes the NC symbol resulting. The destinations apply RS channel decoding before network decoding . In other words, the relays in charge of network coding can also execute channel coding/decoding.
\end{enumerate}
Our simulation results show clearly how scheme 2 outperforms scheme 1.
The network coding operation is performed by the use of a bit-level XOR operator denoted $\oplus$ on the symbols. For example, in $GF(16)$ with 4 bits symbols, $5 \oplus 12 = 9$ and $3 \oplus 5 = 6$. The XOR operation is applied bit per bit in the binary representation of symbols.

RS codes are non-binary cyclic codes with symbols made up of $q-$bit sequences, where $q$ is any positive integer having a value greater than 2. RS($n, k$) codes on $q-$bit symbols exist for all $n$ and $k$ for which
$0 < k < n < 2^q + 2$ where $k$ is the number of data symbols being encoded, and $n$ is the total number of code symbols in the encoded block. For the most conventional RS($n, k$) code,
$(n, k) = (2^q - 1, 2^q - 1 - 2t)$ where $t$ is the symbol-error correcting capability of the code, and $n - k = 2t$ is the number of parity symbols. An extended RS code can be made up with $n = 2^q$ or $n = 2^q + 1$, but not any further \cite{b7}.

Hence, the choice of RS code parameters $n$ and $k$ is not random. Similarly to an optimization problem with constraints, we need to maximize the difference $n-k=2t$, $i.e.$ maximize the number $t$ of corrected symbols while respecting some conditions on $n$ and $k$. These constraints are: $n$ and $k$ must satisfy $0<k<n<2^q+2$ and for our simulations, $k$ must be divisor of the number of symbols in the test sequence.

Our contribution in this paper, in contrast to the previous works dealing with the application of NCC in combination with channel codes to a small network, is the generalization to a wide panoply of larger wireless network topologies using one or two relays and a higher number of source-destination pairs. We also confirm the efficiency of RS channel codes application on finite fields elements $GF(M)$ and the superiority of scheme 2 over scheme 1 in providing best network reliability and performance in terms of symbol error probability.

The wireless network topologies studied in this paper are divided in two categories. The first one uses one relay and the second one uses two in series. In the first one, we simulated the $X-$structure configuration and proposed an extended version containing four sources and four corresponding destinations. With two relays in series, we considered the butterfly network topology and proposed an extended version with four source nodes communicating with four respective destination nodes. All these topologies are illustrated for one source communicating with its destination while the remaining nodes are cooperating and hence increasing the spatial diversity and the overall throughput of the wireless network. We assume that the radio range of each source $S_i$ ($i=1, ..., N$) can reach the relay $R$ and the destinations $D_j$
($j=1, ..., N, j \neq i$).

\subsection{X-structure network}

The first wireless network configuration studied is the well known $X-$topology illustrated in Fig. 1a. In this network, two sources are transmitting messages towards two respective destinations where each source is cooperating to ensure the reliable reception of the other source's message. On the other hand, the relay handles the network coding procedure and broadcasts the result (combination of both messages) to both destinations. This is achieved in 3 time slots while in the same network but without NCC these transmissions would take 4 time slots which lead to a throughput gain of 4/3. The diversity order of the NCC scheme is 2 which can be proved by the following system outage probability calculation. For this network, system outage occurs when the data from source 1 and source 2 cannot both be correctly recovered at their respective destinations \cite{b8}. Let $P_S$ be the system outage probability and let $p_1$, $p_2$ and $p_R$ be the error rates of uplink channels of $S_1$, $S_2$ and the relay R respectively. The expression of $P_S$ is given by

\begin{align*}
P_S&=p_1p_R(1-p_2)+p_2p_R(1-p_1)+p_1p_2(1-p_R)\\ 
& +p_1p_2p_R\\
& =p_1p_2+p_1p_R+p_2p_R-2p_1p_2p_R.
\end{align*}
If we assume similar uplink error rates for all channels transmissions, $i.e.$ $p_1=p_2=p_R=p<<1$, then we have
\begin{equation}\label{}
P_S=3p^2-2p^3 \sim p^2 \sim O\left(\frac{1}{SNR}\right)^2
\end{equation}

Let us first define some useful applications:
\begin{itemize}
\item Let $\mathbb{F}_M$ be the finite field $GF(M)$ to which belong the symbols to transmit. The application $\Gamma : \mathbb{F}_M^k \rightarrow \mathbb{F}_M^n$ that maps $k$ symbols $s_i$ from $\mathbb{F}_M$ into $n$ symbols $x_i$ from $\mathbb{F}_M$ represents our channel encoder RS($n,k$).
\item The application $\Phi : \mathbb{F}_M \rightarrow \mathbb{C}$ that maps a symbol $s$ or $x$ from $\mathbb{F}_M$ into a complex value representing a constellation is our $M-$PSK modulation.
\item The application $\psi : \mathbb{F}_M^N \rightarrow \mathbb{F}_M$ that maps $N$ symbols from $\mathbb{F}_M$ into one is our network coding operation performed in practice by a bit-level XOR between the $N$ symbols, $i.e$, $\psi(x_1,x_2,...,x_N) = x_1\oplus x_2\oplus ... \oplus x_N=\bigoplus_{i=1}^{N} x_i.$
\item The application $\xi : \mathbb{C} \rightarrow \mathbb{C}$ that maps a complex value into another one represents the equalization process at each receiver which consists of multiplying the received signal by the conjugate of the channel fading coefficient.
\end{itemize}

In the absence of NC, we consider two scenarios.
In the first scenario (without RS coding), the message from $S_1$ to $D_1$ is processed as follows:
The symbol $s_1$ is mapped into the $M-$PSK constellation using the application $\Phi$ (PSK modulation) before its transmission to the relay $R$ through a complex AWGN channel with complex fast Rayleigh attenuation. The received symbol at $R$ is $y_{S_1R}=\sqrt{P_1}\Phi(s_1)h_{S_1R}+n_{S_1R}$ where $P_1$ is the transmit power of source 1, $n_{S_1R}$ is the additive white Gaussian noise of the link ($S_1 \rightarrow R$) and $h_{S_1R}$ is the fast fading Rayleigh coefficient (the value of $h_{S_1R}$ varies with each symbol). After reception, the relay first equalizes $y_{S_1R}$ (as we already assumed perfect CSI available) using the equalization operator $\xi$ then simply amplifies and forwards (AF relay) the result. The amplification factor used at the relay is $\beta_{S_1R}$ given by \cite{b9}:
\begin{equation}\label{}
\beta_{S_1R}=\sqrt\frac{P_R}{P_1|h_{S_1R}|^2+\sigma_{S_1R}^2}
\end{equation}
where $\sigma_{S_1R}^2$ is the variance of noise $n_{S_1R}$ on the link ($S_1 \rightarrow R$).
That is, the forwarded symbol from relay $R$ to destination $D_1$ is $y_{RD_1}=\sqrt{P_R}\beta_{S_1R}\xi(y_{S_1R})h_{RD_1}+n_{RD_1}$ where $P_R$ is the relay transmit power, $h_{RD_1}$ and $n_{RD_1}$ are the Rayleigh coefficient and noise component of the link ($R~\rightarrow~D_1$). At destination node $D_1$, the symbol $s_1$ is recovered using the operations $\xi$ and $\Phi$ as follows: $\tilde{s}_1=\Phi^{-1}(\xi(y_{RD_1}))$.

In the second scenario (with RS coding), the symbol received at relay R from $S_1$ is 
\begin{equation}\label{}
y_{S_1R}=\sqrt{P_1}\Phi(\Gamma(s_1))h_{S_1R}+n_{S_1R}.
\end{equation}
The symbol received at destination $D_1$ has the same form as in the first scenario (assuming the relay unable to perform channel coding/decoding) but, of course, with a different value of $y_{S_1R}$, $i.e.$,
\begin{equation}\label{}
y_{RD_1}=\sqrt{P_R}\beta_{S_1R}\xi(y_{S_1R})h_{RD_1}+n_{RD_1}. 
\end{equation}
The destination (now equipped with RS decoder) recovers the transmitted symbol $s_1$ by evaluating 
\begin{equation}\label{}
\tilde{s}_1=\Gamma^{-1}(\Phi^{-1}(\xi(y_{RD_1}))).
\end{equation}
The scenarios above (without NCC) serve as reference to show the improvements in network reliability provided by the application of NCC procedure described below.

When NCC is applied in the \textit{X}-topology (without channel coding application $\Gamma$), both relay $R$ and destination $D_2$ receive the message broadcast from $S_1$ in time slot 1 over the links ($S_1 \rightarrow R$) and ($S_1 \rightarrow D_2$). At $R$, the received signal is $y_{S_1R}=\sqrt{P_1}\Phi(s_1)h_{S_1R}+n_{S_1R}$ and at $D_2$ $y_{S_1D_2}=\sqrt{P_1}\Phi(s_1)h_{S_1D_2}+n_{S_1D_2}$. In time slot 2, $S_2$ sends $s_2$ through both channels ($S_2 \rightarrow R$) and ($S_2 \rightarrow D_1$) and the received signals are, respectively, $y_{S_2R}=\sqrt{P_2}\Phi(s_2)h_{S_2R}+n_{S_2R}$ and $y_{S_2D_1}=\sqrt{P_2}\Phi(s_2)h_{S_2D_1}+n_{S_2D_1}$. The relay demodulates the received symbols using $\Phi^{-1}$ then encodes them using $\psi$, remodulates the result with $\Phi$ and broadcasts the coded symbol to both $D_1$ and $D_2$. Each destination node uses its received symbols to recover its intended message. At $D_1$ for instance, $s_1$ is recovered by 
\begin{equation}\label{}
\tilde{s}_1=\psi^{-1}(\Phi^{-1}(\xi(y_{RD_1})), \Phi^{-1}(\xi(y_{S_2D_1})))
\end{equation}
where 
\begin{align}
y_{RD_1}& =\sqrt{P_R}\Phi(\psi(\Phi^{-1}(\xi(y_{S_1R})),\Phi^{-1}(\xi(y_{S_2R}))))\nonumber\\
       & .h_{RD_1}+n_{RD_1}
\end{align}
and $y_{S_2D_1}=\sqrt{P_2}\Phi(s_2)h_{S_2D_1}+n_{S_2D_1}$.

From now on, we use only $\psi$ for both encoding and decoding operations since we have $\psi^{-1} = \psi$ (the attractive property of XOR).

When we add RS coding/decoding to the scenario above and using scheme 1, (6) becomes
\begin{equation}\label{}
\tilde{s}_1=\Gamma^{-1}(\psi(\Phi^{-1}(\xi(y_{RD_1})), \Phi^{-1}(\xi(y_{S_2D_1}))))
\end{equation}
where
\begin{equation}\label{}
y_{S_2D_1}=\sqrt{P_2}\Phi(\Gamma(s_2))h_{S_2D_1}+n_{S_2D_1}
\end{equation}
and 
\begin{align}
y_{RD_1} & = \sqrt{P_R}\Phi(\psi(\Phi^{-1}(\xi(y_{S_1R})),\Phi^{-1}(\xi(y_{S_2R}))))\nonumber\\ 
         & .h_{RD_1}+n_{RD_1}.
\end{align}

Now, using scheme 2 instead of scheme 1 for the combination of NCC and RS codes, (8) becomes
\begin{equation}\label{}
\tilde{s}_1=\psi(\Gamma^{-1}(\Phi^{-1}(\xi(y_{RD_1}))),\Gamma^{-1}(\Phi^{-1}(\xi(y_{S_2D_1}))))
\end{equation}
where $y_{S_2D_1}$ is described in (9) and
\begin{align}
y_{RD_1} & =\sqrt{P_R}\Phi(\Gamma(\psi(\Gamma^{-1}(\Phi^{-1}(\xi(y_{S_1R}))),\nonumber\\
   & \Gamma^{-1}(\Phi^{-1}(\xi(y_{S_2R}))))))h_{RD_1}+n_{RD_1} 
\end{align}
In the next subsection, we propose an extended version of the \textit{X}-structure dealing with four source-destination pairs instead of two.
\subsection{Extended $X-$structure network}
Fig. 1b shows the proposed extended wireless $X-$network where 4 $(S-D)$ pairs are communicating via one relay. Here, each source $S_i$ broadcasts in the MAC phase the message $s_i$ to the relay $R$ and all destinations $D_j$ ($j\neq i$), $i,j\in \{1,2,3,4\}$. In Fig. 2, we have $i=1$ and the relations obtained in the previous section are extended as follows:
Without NCC, the same relations hold for both scenarios (with and without RS channel coding). When NCC is applied without RS code, 4 time slots are needed to broadcast the 4 signals $s_i$ to the relay and the destinations $D_j$ ($j\neq i$). The received signals at $R$ have the same form
\begin{equation}
y_{S_iR}=\sqrt{P_i}\Phi(s_i)h_{S_iR}+n_{S_iR}, (i=1,\cdots,4)
\end{equation}

The received signal at $D_1$ is the extended version of (6), $i.e.$
\begin{align}
\tilde{s}_1 & =\psi(\Phi^{-1}(\xi(y_{RD_1})), \Phi^{-1}(\xi(y_{S_2D_1})),\nonumber\\
            & \Phi^{-1}(\xi(y_{S_3D_1})),\Phi^{-1}(\xi(y_{S_4D_1})))
\end{align}
where
\begin{align}
y_{RD_1}& =\sqrt{P_R}\Phi(\psi(\Phi^{-1}(\xi(y_{S_1R})),\Phi^{-1}(\xi(y_{S_2R})),\nonumber\\
       & \Phi^{-1}(\xi(y_{S_3R})),\Phi^{-1}(\xi(y_{S_4R}))))h_{RD_1}+n_{RD_1}
\end{align}
that we can write as
\begin{equation*}
y_{RD_1} =\sqrt{P_R}\Phi(\bigoplus_{i=1}^{4}\Phi^{-1}(\xi(y_{S_iR})))h_{RD_1}+n_{RD_1}.
\end{equation*}
Likewise, (14) can be expressed as
\begin{equation*}
\tilde{s}_1 =\Phi^{-1}(\xi(y_{RD_1}))\oplus\bigoplus_{i=2}^{4}\Phi^{-1}(\xi(y_{S_iD_1})).
\end{equation*}
For all scenarios, the results are the same but with 4  $(S-D)$ pairs instead of 2. For instance, with NCC and RS codes in scheme 2, (11) becomes
\begin{align}
\tilde{s}_1&=\psi(\Gamma^{-1}(\Phi^{-1}(\xi(y_{RD_1}))),\Gamma^{-1}(\Phi^{-1}(\xi(y_{S_2D_1}))),\nonumber\\
 & \Gamma^{-1}(\Phi^{-1}(\xi(y_{S_3D_1}))),\Gamma^{-1}(\Phi^{-1}(\xi(y_{S_4D_1}))))
\end{align}


\subsection{Butterfly network}
In the butterfly wireless network of Fig. 1c, there are two source-destination ($S-D$) pairs communicating via 2 relays in series. The first relay $R_1$ is the one in charge of network coding and channel coding/decoding while the second $R_2$ simply forwards (AF protocol) the received signal from $R_1$. Then, destinations $D_1$ and $D_2$ receive the broadcast message from $R_2$.
In the conventional scenario (without NCC), a transmission between both $(S-D)$ pairs requires 6 time slots while with NCC, we need only 4 which leads to a throughput gain of $3/2$.   

In the direct path scenario with no NCC and no RS codes, the symbol received at $R_1$ is $y_{S_1R_1}=\sqrt{P_1}\Phi(s_1)h_{S_1R_1}+n_{S_1R_1}$. Relay $R_2$ receives $y_{R_1R_2}=\sqrt{P_{R_1}}\beta_{S_1R_1}\xi(y_{S_1R_1})h_{R_1R_2}+n_{R_1R_2}$ from $R_1$ where
\begin{equation}
\beta_{S_1R_1}=\sqrt\frac{P_{R_1}}{P_1|h_{S_1R_1}|^2+\sigma_{S_1R_1}^2}
\end{equation}
and finally, destination $D_1$  receives $y_{R_2D_1}=\sqrt{P_{R_2}}\beta_{R_1R_2}\xi(y_{R_1R_2})h_{R_2D_1}+n_{R_2D_1}$ from relay $R_2$ where 
\begin{equation}
\beta_{R_1R_2}=\sqrt\frac{P_{R_2}}{P_{R_1}|h_{R_1R_2}|^2+\sigma_{R_1R_2}^2}
\end{equation}
and retrieves its intended symbol $\tilde{s}_1=\Phi^{-1}(\xi(y_{R_2D_1}))$.

Considering the application of channel coding at both sources and channel decoding at their respective destinations for the previous scenario, we have $y_{S_1R_1}=\sqrt{P_1}\Phi(\Gamma(s_1))h_{S_1R_1}+n_{S_1R_1}$
and $\tilde{s}_1=\Gamma^{-1}(\Phi^{-1}(\xi(y_{R_2D_1})))$.
When NCC is performed without RS channel coding, the relay $R_2$ receives from $R_1$ the signal 
\begin{equation}
y_{R_1R_2}=\sqrt{P_{R_1}}\Phi(\bigoplus_{i=1}^{2}\Phi^{-1}(\xi(y_{S_iR_1})))h_{R_1R_2}+n_{R_1R_2}.
\end{equation}
The retrieved symbol at $D_1$ is
\begin{equation}
\tilde{s}_1 =\Phi^{-1}(\xi(y_{R_2D_1}))\oplus\Phi^{-1}(\xi(y_{S_2D_1})).
\end{equation}
Combining RS codes with NCC respecting scheme 1 leads to the following relations:
$y_{R_1R_2}$ is given by (19) with the difference that received symbols at $R_1$ are now RS coded and have the form
\begin{equation*}
y_{S_iR_1}=\sqrt{P_i}\Phi(\Gamma(s_i))h_{S_iR_1}+n_{S_iR_1}.
\end{equation*}
In this case, destination $D_1$ recovers the symbol $s_1$ with the operation
\begin{align}
\tilde{s}_1&=\Gamma^{-1}(\psi(\Phi^{-1}(\xi(y_{R_2D_1})), \Phi^{-1}(\xi(y_{S_2D_1}))))\nonumber\\
           &=\Gamma^{-1}(\Phi^{-1}(\xi(y_{R_2D_1}))\oplus\Phi^{-1}(\xi(y_{S_2D_1}))).
\end{align}
Now, combining RS codes with NCC respecting scheme 2, we have
\begin{align}
y_{R_1R_2} & =\sqrt{P_{R_1}}\Phi(\Gamma(\psi(\Gamma^{-1}(\Phi^{-1}(\xi(y_{S_1R_1}))),\nonumber\\
   & \Gamma^{-1}(\Phi^{-1}(\xi(y_{S_2R_1}))))))h_{R_1R_2}+n_{R_1R_2}. 
\end{align}
After equalization of $y_{R_1R_2}$, the relay $R_2$ amplifies it using $\beta_{R_1R_2}$ and broadcasts it to $D_1$ and $D_2$. Hence, $D_1$ decodes the symbol $s_1$ as follows:
\begin{equation}\label{}
\tilde{s}_1=\psi(\Gamma^{-1}(\Phi^{-1}(\xi(y_{R_2D_1}))),\Gamma^{-1}(\Phi^{-1}(\xi(y_{S_2D_1})))).
\end{equation}

\subsection{Extended Butterfly network}

In this subsection, we propose to extend the number of $(S-D)$ pairs to 4 as illustrated in Fig. 1d. The results found for the butterfly network hold for its extended version with the addition of sources $S_3$ and $S_4$ and destinations $D_3$ and $D_4$. In the scenario of scheme 1 for example, (21) becomes
\begin{align}
\tilde{s}_1&=\Gamma^{-1}(\psi(\Phi^{-1}(\xi(y_{R_2D_1})), \Phi^{-1}(\xi(y_{S_2D_1})),\nonumber\\
           &\Phi^{-1}(\xi(y_{S_3D_1})), \Phi^{-1}(\xi(y_{S_4D_1})))).
\end{align}
that can be expressed as
\begin{equation}
\tilde{s}_1=\Gamma^{-1}(\Phi^{-1}(\xi(y_{R_2D_1})) \oplus \bigoplus_{i=2}^{4}\Phi^{-1}(\xi(y_{S_iD_1}))).
\end{equation}
Application of NCC on the extended butterfly network allows a throughput gain of $2$ with the diversity order 4. These values become $3N/(N+2)$ and $N$ for the general case of $N$ $(S-D)$ pairs.

In the next section, we get the results that confirm the efficiency of combining NCC with RS channel codes by conducting simulations of all the scenarios above. 

\section{Simulation Results}
The scenarios detailed in the wireless topologies presented in the previous section are simulated using MATLAB. We generate a set of 14 SNR values from $0 dB$ to $26 dB$ with a step of $2 dB$. For each $M-$PSK constellation, we generate (at each source node) a sequence of 1000 random integer values in the corresponding finite field $GF(M)$. Depending on the coding rate $k/n$ of our RS coder, these values are coded into a set of $\frac{1000n}{k}$ values in the same finite field. We assume equal transmit power at all nodes. All channels are independent and each one is defined by its complex AWGN components $n  \sim CN(0, \sigma_n^2)$ and complex Rayleigh fading coefficients $h \sim CN(0, \sigma_h^2)$. Each simulation is the average of 1000 iterations to ensure accurate reliable results.

To derive the values of SER, we compare the received sequence of symbols $\tilde{s_1}$ evaluated using the corresponding scenario equation to the original one $s_1$ generated at source node $S_1$. 

\begin{figure}[!t]
\centerline{\includegraphics[scale=0.65]{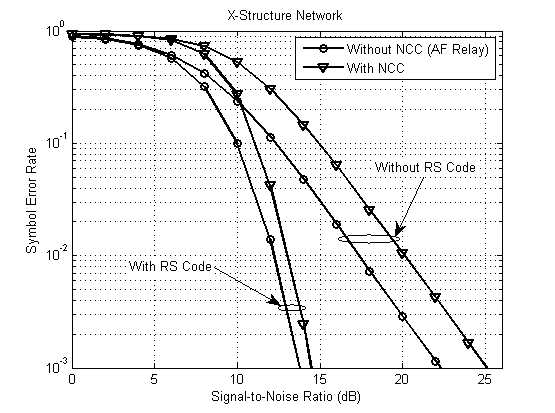}}
\caption{SER vs. SNR obtained for $X$-structure with 16-PSK and RS(15,5).}
\label{fig}
\end{figure}



Due to the large number of scenarios in our work, we gathered the main relevant results in one SER versus SNR graph for each topology in addition to one graph ($X-$structure with $16-$PSK modulation and RS(15,5)) to illustrate the advantage of NCC application with and without Reed-Solomon channel coding. In Fig. 2, we can notice how the application of cooperative network coding, at the cost of a small degradation of SER values, produces an important improvement of the overall system throughput and diversity order \cite{b10} in addition to transmit power reduction. Table I shows the values of throughput gain and diversity orders attained by exploiting NCC technique on the general cases of the networks studied in this paper. Also, RS channel coding (we used scheme 2) allows considerable reductions of symbol error probability. At $SNR=14.35dB$, the SER with RS code is equal to $10^{-3}$ while its value is $10^{-0.9}$ without RS code. For all the four wireless network configurations, the graphs on Figs. 3 and 4 show clearly how scheme 2 outperforms scheme 1 in terms of symbol error probability for the three PSK constellations adopted. For instance, in the extended butterfly network using $32-$PSK and RS(31,10), the SER decreases from $10^{-1}$ in scheme 1 to $10^{-3}$ in scheme 2 at $SNR=16.65dB$. Furthermore, with a small extra computation at the relay in scheme 2 to perform $RS(n,k)$ codes of the received messages and re-encoding of the resulting network coded message, we can save a large amount of system transmit energy \cite{b11,b12,b13}, since, in scheme 1, the number of NC symbols to transmit is $\frac{n}{k}$ times greater than the number in scheme 2. 

\begin{table}[!t]
\label{}
\caption{Throughput gain and Diversity Order.}
\begin{center}
\begin{tabular}{| c | c | c | c | c |}
\hline
\textbf{} & \multicolumn{4}{ c |}{\textbf{Network Configuration}}  \\ 
\cline{2-5}
& \textbf{X-net.} & \textbf{Extended-X} & \textbf{Butter.} & \textbf{Ext. Butter.} \\
\hline

Throughput Gain & $\frac{4}{3}$ & $\frac{2N}{N+1}$ & $\frac{3}{2}$ & $\frac{3N}{N+2}$ \\ \hline
Diversity Order & $2$ &  $N$ & $2$  & $N$\\ \hline

\end{tabular}
\end{center}
\end{table}

\begin{figure}[!h]
\centerline{\includegraphics[scale=0.34]{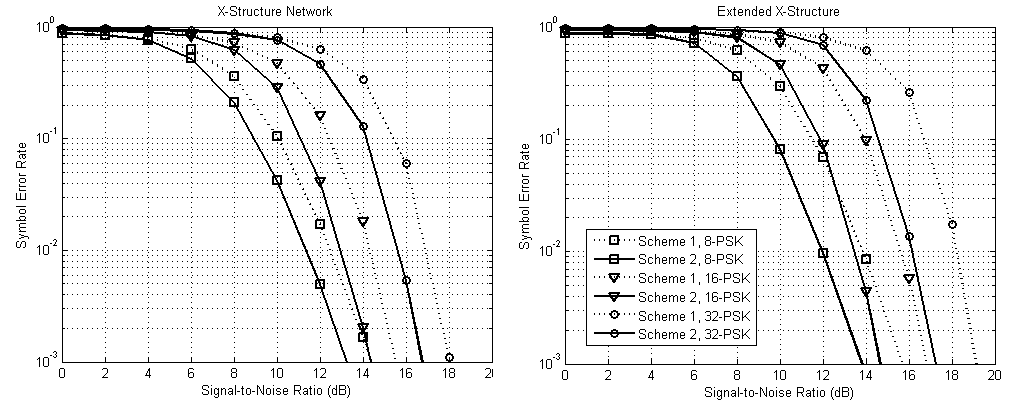}}
\caption{Results of scheme 1 and scheme 2 on X-structure and extended X-structure networks with RS(7,2), RS(15,5) and RS(31,10) for 8-PSK, 16-PSK and 32-PSK respectively.}
\label{fig}
\end{figure}

\begin{figure}[!h]
\centerline{\includegraphics[scale=0.34]{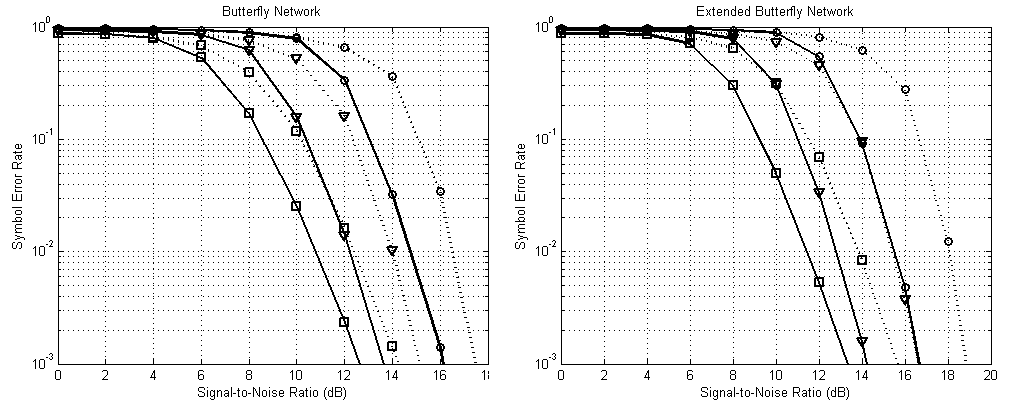}}
\caption{Results of scheme 1 and scheme 2 on butterfly and extended butterfly networks with RS(7,2), RS(15,5) and RS(31,10) for 8-PSK, 16-PSK and 32-PSK respectively.}
\label{fig}
\end{figure}

\section{Conclusion}

In this paper, we put the light on the advantage of using Reed-Solomon codes jointly with cooperative network coding in wireless network architectures for non-binary PSK constellations. Besides the optimizations achieved in terms of system throughput, diversity order and network transmit power by NCC strategy, we showed how RS channel codes can improve considerably the symbol error rate. RS codes give high performance with non-binary constellations since they deal with Galois fields $GF(M)$ elements. We also proved that scheme 2, where the relay operates RS decoding before network coding and then RS re-encoding, decreases the symbol error probability in comparison with scheme 1 where the relay executes network coding directly on channel coded symbols. As a future work, we will investigate the application of random linear network coding (RLNC) at the NC relay.

\end{document}